\documentclass[a4paper]{article}
\usepackage{hyperref}
\usepackage{algorithm}
\usepackage[noend]{algpseudocode}
\usepackage{enumitem}

\makeatletter
\def\BState{\State\hskip-\ALG@thistlm}
\makeatother

\usepackage{INTERSPEECH2018}

\title{Singing voice phoneme segmentation by hierarchically inferring syllable and phoneme onset positions}
\name{Rong Gong, Xavier Serra}
\address{Music Technology Group, Universitat Pompeu Fabra, Barcelona, Spain}
\email{rong.gong@upf.edu, xavier.serra@upf.edu}

\begin{document}

\maketitle
\begin{abstract}
In this paper, we tackle the singing voice phoneme segmentation problem in the singing training scenario by using language-independent information -- onset and prior coarse duration. We propose a two-step method.  In the first step, we jointly calculate the syllable and phoneme onset detection functions (ODFs) using a convolutional neural network (CNN). In the second step, the syllable and phoneme boundaries and labels are inferred hierarchically by using a duration-informed hidden Markov model (HMM). To achieve the inference, we incorporate the \textit{a priori} duration model as the transition probabilities and the ODFs as the emission probabilities into the HMM. The proposed method is designed in a language-independent way such that no phoneme class labels are used. For the model training and algorithm evaluation, we collect a new jingju (also known as Beijing or Peking opera) solo singing voice dataset and manually annotate the boundaries and labels at phrase, syllable and phoneme levels. The dataset is publicly available. The proposed method is compared with a baseline method based on hidden semi-Markov model (HSMM) forced alignment. The evaluation results show that the proposed method outperforms the baseline by a large margin regarding both segmentation and onset detection tasks.
  
\end{abstract}

\section{Introduction}

\subsection{Background}
\label{sec:background}
The objective of this work lies in the background of automatic singing voice pronunciation assessment (ASVPA) for jingju music. Jingju singing is sung in the Mandarin language varied by two Chinese dialects. In a professional jingju singing training situation, the teacher would be very demanding of the students regarding a precise pronunciation of each syllable and phoneme. 

We design an ASVPA system according to the \textit{learning by imitation} method which is used as the basic training method by many musical traditions and as well by jingju singing \cite{gong2017identification}. In practice, this method contains three steps in regards to teaching how to sing a musical phrase: (i) the teacher firstly gives a demonstrative singing, (ii) Then the student is asked to imitate it. (iii) The teacher provides the feedback by assessing the student's singing at the syllable or phoneme-levels. Steps (ii) and (iii) should be repeated until the teacher satisfies with the student's singing. We design a three-steps ASVPA system in consideration of the above training method: (a) the teacher's demonstrative and student's imitative singing voice audios are recorded. The former is manually pre-segmented and labeled at the phrase, syllable and phoneme-levels. The latter is manually pre-segmented and labeled only at phrase-level. (b) The student's singing voice is then automatically segmented and labeled at the phoneme-level using the proposed method in this paper. (c) The corresponding phonemes between teacher's and student's recordings are finally compared by a phonetic pronunciation similarity algorithm. The similarity score will be given to the student as her/his pronunciation score.

In this paper, we approach step (a) manually so that the coarse durations and labels of the teacher's demonstrative singing are available as the prior information for step (b). We tackle the phoneme segmentation problem of step (b). Step (c) remains a work in progress.

\subsection{Related work}

The first topic which is highly related to our research is speech forced alignment. Speech forced alignment is a process that the orthographic transcription is aligned with the speech audio at word or phone-level. Most of the non-commercial alignment tools are built on HTK \cite{Young2006HTK} or Kaldi \cite{Povey2011ASRU} frameworks, such as Montreal forced aligner \cite{McAuliffe2017} and Penn Forced Aligner \cite{PennForced}. These tools implement a part of the automatic speech recognition (ASR) pipeline, train the HMM acoustic models iteratively using Viterbi algorithm and align audio features (e.g. MFCCs) to the HMM states. Brognaux and Drugman \cite{brognaux2016hmm} explored the forced alignment on a small dataset using supplementary acoustic features and initializing the silence model by voice activity detection algorithm. To predict the confidence measure of the aligned word boundaries and to fine-tune their time positions, Serri\'{e}re et al. \cite{serriere2016weakly} explored an alignment post-processing method using a deep neural network (DNN). The forced alignment is language-dependent, in which the acoustic models should be trained by using the corpus of a certain language. Another category of speech segmentation methods is language-independent, which relies on detecting the phoneme boundary change in the temporal-spectral domain \cite{esposito2005text,almpanidis2009Robust}. The drawback of these methods is that the segmentation accuracies are poorer than the language-dependent counterparts \cite{pakoci2016phonetic}. 

The second topic related to our research is the singing voice lyrics-to-audio alignment. Most of these works \cite{mesaros2008automatic, loscos1999Low, fujihara2011lyricsynchronizer, mauch2012integrating, iskandar2006syllabic,gong2015real, kruspe2015keyword, dzhambazov2015modeling} used the forced alignment method accompanied by music-related techniques. Loscos et al. \cite{loscos1999Low} used MFCCs with additional features and also explored specific HMM topologies. Fujihara et al. \cite{fujihara2011lyricsynchronizer} used voice/accompaniment separation to deal with mixed recording, and vocal detection, fricative detection to increase the alignment performance. Additional musical side information extracted from the musical score is used in many works. Mauch et al. \cite{mauch2012integrating} used chord information such that each HMM state contains both chord and phoneme labels. Iskandar et al. \cite{iskandar2006syllabic} constrained the alignment by using musical note length distribution. Gong et al. \cite{gong2015real}, Kruspe \cite{kruspe2015keyword}, Dzhambazov and Serra \cite{dzhambazov2015modeling} all used syllable/phoneme duration extracted from the musical score and decoded the alignment path by duration-explicit HMM models. Chien et al. \cite{Chien2016Alignment} introduced an approach based on vowel likelihood models. Chang and Lee \cite{Chang2017Lyrics} used canonical time warping and repetitive vowel patterns to find the alignment for vowel sequence. Some other works achieved the alignment at music structure-level \cite{muller2007lyrics} or line-level \cite{wang2004lyrically}.

Our research is also related to multi-task learning (MTL) because we want to achieve the segmentation on the jointly learned syllable and phoneme ODFs. MTL means that learning by optimizing more than one loss function \cite{ruder2017overview,caruana1998multitask}. Hard parameter sharing is the most commonly used MTL approach, which applied by sharing the hidden layers between all tasks and keep several task-specific output layers \cite{ruder2017overview}. Baxter argued that hard parameter sharing can reduce the risk of overfitting in an order of the number of tasks. In the music information retrieval (MIR) domain, Yang et al. proposed an MTL framework based on the neural networks to jointly consider chord and root note recognition problems \cite{Yang2016highlighting}. Vogl et al. showed that learning beats jointly with drums can be beneficial for the task of drum detection \cite{Vogl2017DrumTV}.

\subsection{Contribution}
In this paper, we present a new jingju solo singing voice dataset for the phoneme segmentation, which is manually annotated at three hierarchical levels - phrase, syllable, phoneme (section \ref{sec:dataset}). We propose a language-independent phoneme segmentation method which jointly learns syllable and phoneme onsets and hierarchically infers the phoneme boundaries and labels by a duration-informed HMM (section \ref{sec:proposed}). Finally, we build a forced alignment baseline method based on HSMM and compare it with the proposed method (section \ref{sec:eval}).

\section{Dataset}
\label{sec:dataset}
The jingju solo singing voice dataset focuses on two most important jingju role-types (performing profile) \cite{repetto_creating_2014}: \textit{dan} (female) and \textit{laosheng} (old man). It has been collected by the researchers in Centre for Digital Music, Queen Mary University of London \cite{black_automatic_2014} and Music Technology Group, Universitat Pompeu Fabra. 
\begin{table}[ht]
	\centering
	\caption{Statistics of the dataset}
	\label{table:detailInfoDataset}
    \resizebox{\columnwidth}{!}{
	\begin{tabular}{l|cccc}
		\toprule
		& \#Recordings & \#Phrases & \#Syllables & \#Phonemes \\
		\midrule
		Train           & 56 & 214 & 1965 & 5017 \\
		Test   			& 39 & 216 & 1758 & 4651  \\
		\bottomrule
	\end{tabular}
    }
\end{table}

The dataset contains 95 recordings split into train and test sets (table \ref{table:detailInfoDataset}). The recordings in the test set are student imitative singing. Their teacher demonstrative recordings can be found in the train set, which guarantees that the coarse syllable/phoneme duration and labels are available for the algorithm testing. Audios are pre-segmented into singing phrases. The syllable/phoneme ground truth boundaries (onsets/offsets) and labels are manually annotated in Praat \cite{boersma_praat_2001} by two Mandarin native speakers and a jingju musicologist. 29 phoneme categories are annotated, which include a silence category and a non-identifiable phoneme category, e.g. throat-cleaning. The category table can be found in the Github page\footnote{\label{ft:github}\url{https://goo.gl/fFr9XU}}. The dataset is publicly available\footnote{\url{https://doi.org/10.5281/zenodo.1185123}}.

\section{Proposed method}
\label{sec:proposed}
We introduce a coarse duration-informed phoneme segmentation method. The syllable and phoneme onset ODFs are jointly learned by a hard parameter sharing multi-task CNN model. The syllable/phoneme boundaries and labels are then inferred by an HMM using the \textit{a priori} duration model as the transition probabilities and the ODFs as the emission probabilities.

\subsection{CNN onset detection function}
\label{sec:cnn_odf}
\noindent\textbf{Audio preprocessing}: We use \textsc{Madmom}\footnote{\url{https://github.com/CPJKU/madmom}} Python package to calculate the log-mel spectrogram of the student's singing audio. The frame size and hop size of the spectrogram are respectively 46.4ms (2048 samples) and 10ms (441 samples). The low and high frequency bounds are 27.5Hz and 16kHz. We use a log-mel context as the CNN model input, where the context size is 80$\times$15 (log-mel bins×frames). Thus the CNN model takes a binary onset/non-onset decision sequentially for every frame given its context: $\pm$70ms, 15 frames in total.

\noindent\textbf{Preparing target labels}: The target labels of the training set are prepared according to the ground truth annotations. We set the label of a certain context to 1 if an onset has been annotated for its corresponding frame, otherwise 0. To compensate the human annotation inaccuracy and to augment the positive sample size, we also set the labels of the two neighbor contexts to 1. However, the importance of the neighbor contexts should not be equal to their center context, thus we compensate this by setting the sample weights of the neighbor contexts to 0.25. A similar sample weighting strategy has been presented in Schluter's paper \cite{schluter2014improved}. Finally, for each log-mel context, we have its syllable and phoneme labels. They will be used as the training targets in the CNN model to predict the onset presence.
\begin{figure}[ht!]
    \centering
    \includegraphics[width=0.5\textwidth]{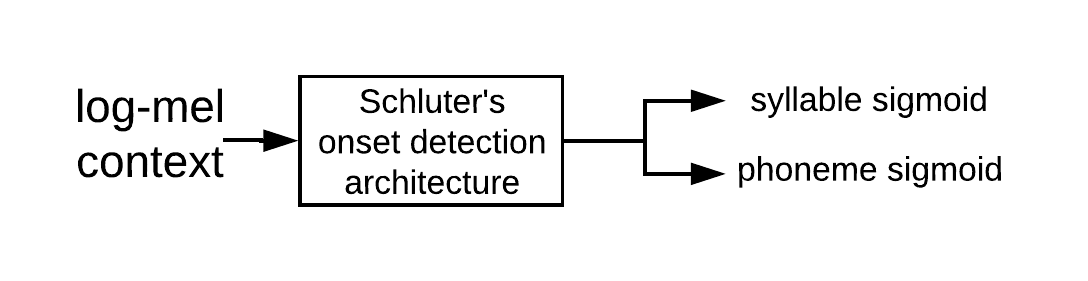}
    \caption{Diagram of the multi-task CNN model.}
    \label{fig:cnn_architecture}
\end{figure}

\noindent\textbf{Hard parameter sharing multi-task CNN model}: We build a CNN for classifying each log-mel context and output the syllable and phoneme ODFs. We extend the CNN architecture presented in Schluter's work \cite{schluter2014improved} by using two predicting objectives -- syllable and phoneme (figure \ref{fig:cnn_architecture}). The two objectives share the same parameters, and both are using the sigmoid activation function. Binary cross-entropy is used as the loss function. The loss weighting coefficients for the two objectives are set to equal since no significant effect has been found in the preliminary experiment. The model parameters are learned with mini-batch training (batch size 256), adam \cite{kingma2014adam} update rule and early stopping -- if validation loss is not decreasing after 15 epochs. The ODFs output from the CNN model is used as the emission probabilities for the syllable/phoneme boundary inference.

\subsection{Phoneme boundaries and labels inference}
\label{sec:dur_label}
The inference algorithm receives the syllable and phoneme durations and labels of teacher's singing phrase as the prior input and infers the syllable and phoneme boundaries and labels for the student's singing phrase.
\subsubsection{Coarse duration and \textit{a priori} duration model}
\label{sec:pp_coarse_duration}
\begin{figure}[ht!]
    \centering
    \includegraphics[width=0.48\textwidth]{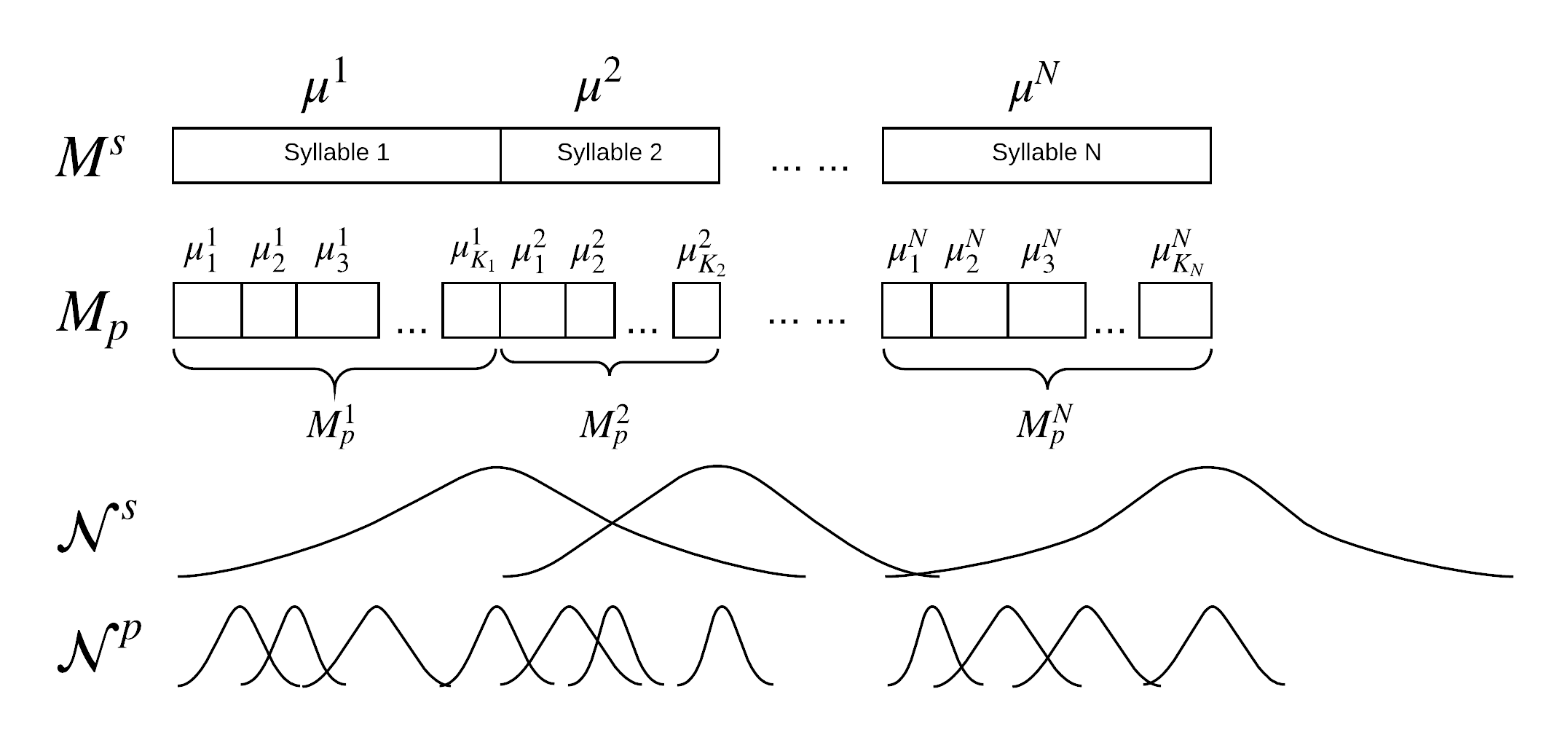}
    \caption{Illustration of the syllable $M^s$ and phoneme $M_p$ coarse duration sequences and their \textit{a priori} duration models -- $\mathcal{N}^s$, $\mathcal{N}^p$. The blank rectangulars in $M_p$ represent the phonemes.}
    \label{fig:coarse_dur}
\end{figure}

The syllable durations of the teacher's singing phrase are stored in an array $M^s=\mu^{1} \cdots \mu^{n} \cdots \mu^{N}$, where $\mu^{n}$ is the duration of the nth syllable. The phoneme durations are stored in a nested array $M_p=M^{1}_p \cdots M^{n}_p \cdots M^{N}_p$, where $M^{n}_p$ is the sub-array with respect to the nth syllable and can be further expanded to $M^{n}_p=\mu_{1}^{n} \cdots \mu_{k}^{n} \cdots \mu_{K_{n}}^{n}$, where $K_{n}$ is the number of phonemes contained in the nth syllable. The phoneme durations of the nth syllable sum to its syllable duration: $\mu^{n}=\sum_{k=1}^{K_{n}} \mu_k^{n}$ (figure \ref{fig:coarse_dur}). In both syllable and phoneme duration sequences -- $M^s$, $M_p$, the duration of the silence is not treated separately and is merged with its previous syllable or phoneme.

The \textit{a priori} duration model is shaped with a Gaussian function $\mathcal{N} (d; {\mu}_n, \sigma_n^2)$. It provides the prior likelihood of an onset to occur according to the syllable/phoneme duration of the teacher's singing. The mean ${\mu}_n$ of the Gaussian represents the expected duration of nth teacher's syllable/phoneme. Its standard deviation $\sigma_n$ is proportional to $\mu_n$: $\sigma_n=\gamma \mu_n$ and $\gamma$ is heuristically set to 0.35. Figure \ref{fig:coarse_dur} provides an intuitive example of how the \textit{a priori} duration model works. The a priori duration model will be incorporated into a duration-informed HMM as the state transition probabilities to inform that where syllable/phoneme onsets is likely to occur in student's singing phrase.

\subsubsection{Duration-informed HMM for segment boundary and label inference}
We present an HMM configuration which makes use of the coarse duration and label input (section \ref{sec:pp_coarse_duration}) and can be applied to inferring firstly (i) the syllable boundaries and labels on the ODF for the whole singing phrase, then (ii) the phoneme boundaries and labels on the ODF segment constrained by the inferred syllable boundaries. To use the same inference formulation, we unify the notations $N$, $K_n$ (both introduced in section \ref{sec:pp_coarse_duration}) to $N$, and $M^s$, $M^{n}_{p}$ to $M$. The unification of the notations has a practical meaning because we use the same algorithm for both syllable and phoneme inference. The HMM is characterized by the following:
\begin{enumerate}[leftmargin=*]
    \item The hidden state space is a set of $T$ candidate onset positions $S_1, S_2, \cdots, S_T$ discretized by the hop size, where $S_{T}$ is the offset position of the last syllable or the last phoneme within a syllable.
    \item The state transition probability at the time instant $t$ associated with state changes is defined by \textit{a priori} duration distribution $\mathcal{N} (d_{ij} ; \mu_t, \sigma_t^2)$, where $d_{ij}$ is the time distance between states $S_i$ and $S_j$ ($j>i$). The length of the inferred state sequence is equal to $N$.
    \item The emission probability for the state $S_j$ is represented by its value in the ODF, which is denoted as $p_j$.
\end{enumerate} 

The goal is to find the best onset state sequence $Q={q_1 q_2 \cdots q_{N-1}}$ for a given duration sequence $M$ and impose the corresponding segment label, where $q_i$ denotes the onset of the $i+1$th inferred syllable/phoneme. The onset of the current segment is assigned as the offset of the previous segment. $q_0$ and $q_N$ are fixed as $S_1$ and $S_T$ as we expect that the onset of the first syllable(or phoneme) is located at the beginning of the singing phrase(or syllable) and the offset of the last syllable(or phoneme) is located at the end of the phrase(or syllable). One can fulfill this assumption by truncating the silences at both ends of the incoming audio. The best onset sequence can be inferred by the logarithmic form of Viterbi algorithm \cite{rabiner1989tutorial}:
% \begin{equation}
% \delta_n(i)= \max_{q_1,q_2,\cdots,q_n}{\log P[q_1 q_2 \cdots q_n,\, \mu_1 \mu_2 \cdots \mu_n]}
% \end{equation}

\begin{algorithm}
\caption{Logarithmic form of Viterbi algorithm using the \textit{a priori} duration model}\label{alg:log_viterbi}
\begin{algorithmic}
\item $\delta_n(i) \gets \max\limits_{q_1,q_2,\cdots,q_n}{\log P[q_1 q_2 \cdots q_n,\, \mu_1 \mu_2 \cdots \mu_n]}$
\Procedure{LogFormViterbi}{$M, p$}

\BState \emph{Initialization}:
\State $\delta_1(i) \gets \log(\mathcal{N} (d_{1i} ; \mu_1, \sigma_1^2))+\log(p_i)$
\State $\psi_1(i) \gets S_1$

\BState \emph{Recursion}:
\State $tmp\_var(i,j) \gets \delta_{n-1} (i) + \log(\mathcal{N} (d_{ij} ; \mu_n, \sigma_n^2))$
\State $\delta_n (j) \gets \max\limits_{1 \leqslant i < j} tmp\_var(i,j) +\log(p_j)$
\State $\psi_n (j) \gets \arg\max\limits_{1 \leqslant i < j} tmp\_var(i,j)$
\BState \emph{Termination}:
% \State $tmp\_var(i) \gets \delta_{N-1} (i) + \log(\mathcal{N} (d_{i T} ; \mu_N, \sigma_N^2))$
% \State $\log P^* \gets \max\limits_{1 \leqslant i < T} tmp\_var(i)$
\State $q_{N} \gets \arg\max\limits_{1 \leqslant i < T} {\delta_{N-1} (i) + \log(\mathcal{N} (d_{i T} ; \mu_N, \sigma_N^2))}$
\EndProcedure
\end{algorithmic}
\end{algorithm}

Finally, the state sequence $Q$ is obtained by the backtracking step.  The implementation of the algorithm can be found in the Github link\textsuperscript{\ref{ft:github}}.

\section{Evaluation}
\label{sec:eval}
The phoneme segmentation task consists of determining the time positions of phoneme onsets and offsets and its labels. As the onset of the current phoneme is assigned as the offset of the previous phoneme, the evaluation consists in comparing the detected (i) onsets and (ii) segments to their reference ones. As the syllable segmentation is the prerequisite for the proposed method, we also report the syllable segmentation results. To compare with the proposed method, we firstly introduce a forced alignment-based baseline method.

\subsection{Baseline method}
\label{sec:baseline}
The baseline is a 1-state monophone DNN/HSMM model. We use monophone model because our small dataset doesn't have enough phoneme instances for exploring the context-dependent triphones model, also Brognaux and Drugman \cite{brognaux2016hmm} and Pakoci et al. \cite{pakoci2016phonetic} argued that context-dependent model can't bring significant alignment improvement. It is convenient to apply 1-state model because each phoneme can be represented by a semi-Markovian state carrying a state occupancy time distribution. The audio preprocessing step is the same as in section \ref{sec:cnn_odf}.

We construct an HSMM for phoneme segment inference. The topology is a left-to-right semi-Markov chain, where the states represent sequentially the phonemes of the teacher's singing phrase. As we are dealing with the forced alignment, we constraint that the inference can only be started by the leftmost state and terminated to the rightmost state. The self-transition probabilities are set to 0 because the state occupancy depends on the predefined distribution. Other transitions -- from current states to subsequent states are set to 1. We use a one-layer CNN with multi-filter shapes as the acoustic model \cite{Pons2017Timbre} and the Gaussian $\mathcal{N} (d; {\mu_n}, \sigma_n^2)$ introduced in section \ref{sec:pp_coarse_duration} as the state occupancy distribution. The inference goal is to find best state sequence, and we use Gu\'{e}don's HSMM Viterbi algorithm \cite{Guedon2017Exploring} for this purpose. The baseline details and code can be found in the Github page\textsuperscript{\ref{ft:github}}. Finally, the segments are labeled by the alignment path, and the phoneme onsets are taken on the state transition time positions.

\subsection{Metrics and results}
To define a correctly detected onset, we choose a tolerance of $\tau=25$ms. If the detected onset $o_d$ lies within the tolerance of its ground truth counterpart $o_g$: $|o_d-o_g|<\tau$, we consider that it's correctly detected. To measure the segmentation correctness, we use the ratio between the duration of correctly labeled segments and the total duration of the singing phrase. This metric has been suggested by Fujihara et al. \cite{fujihara2011lyricsynchronizer} in their lyrics alignment work. We trained both proposed and baseline models 5 times with different random seeds, and report the mean and the standard deviation score on the test set. Below is the evaluation results table:

\begin{table}[ht]
	\centering
	\caption{Evaluation results table. Table cell: mean score$\pm$standard deviation score.}
	\label{table:final_results}
	\begin{tabular}{l|cccc}
		\toprule
        & \multicolumn{2}{c}{Onset F1-measure \%} & \multicolumn{2}{c}{Segmentation \%} \\
		& phoneme & syllable &  phoneme & syllable \\
		\midrule
		Proposed           & 75.2$\pm$0.6 & 75.8$\pm$0.4 & 60.7$\pm$0.4 & 84.6$\pm$0.3 \\
		Baseline   		   & 44.5$\pm$0.9 & 41.0$\pm$1.0 & 53.4$\pm$0.9 & 65.8$\pm$0.7 \\
		\bottomrule
	\end{tabular}
\end{table}

We only show the F1-measure of the onset detection results in table \ref{table:final_results}. The full results can be found in the Github page\textsuperscript{\ref{ft:github}}. 

\subsection{Discussions}

% Why choose proposed method for the singing learning scenario?
On both metrics -- onset detection and segmentation, the proposed method outperforms the baseline. The proposed method uses the ODF which provides the time ``anchors'' for the onset detection. Besides, the ODF calculation is a binary classification task. Thus the training data for both positive and negative class is more than abundant. Whereas, the phonetic classification is a harder task because many singing interpretations of different phonemes have the similar temporal-spectral patterns. Our relatively small training dataset might be not sufficient to train a proper discriminative acoustic model with 29 phoneme categories. We believe that these reasons lead to a better onset detection and segmentation performance of the proposed method.

\begin{figure}[ht!]
    \centering
    \includegraphics[width=0.45\textwidth]{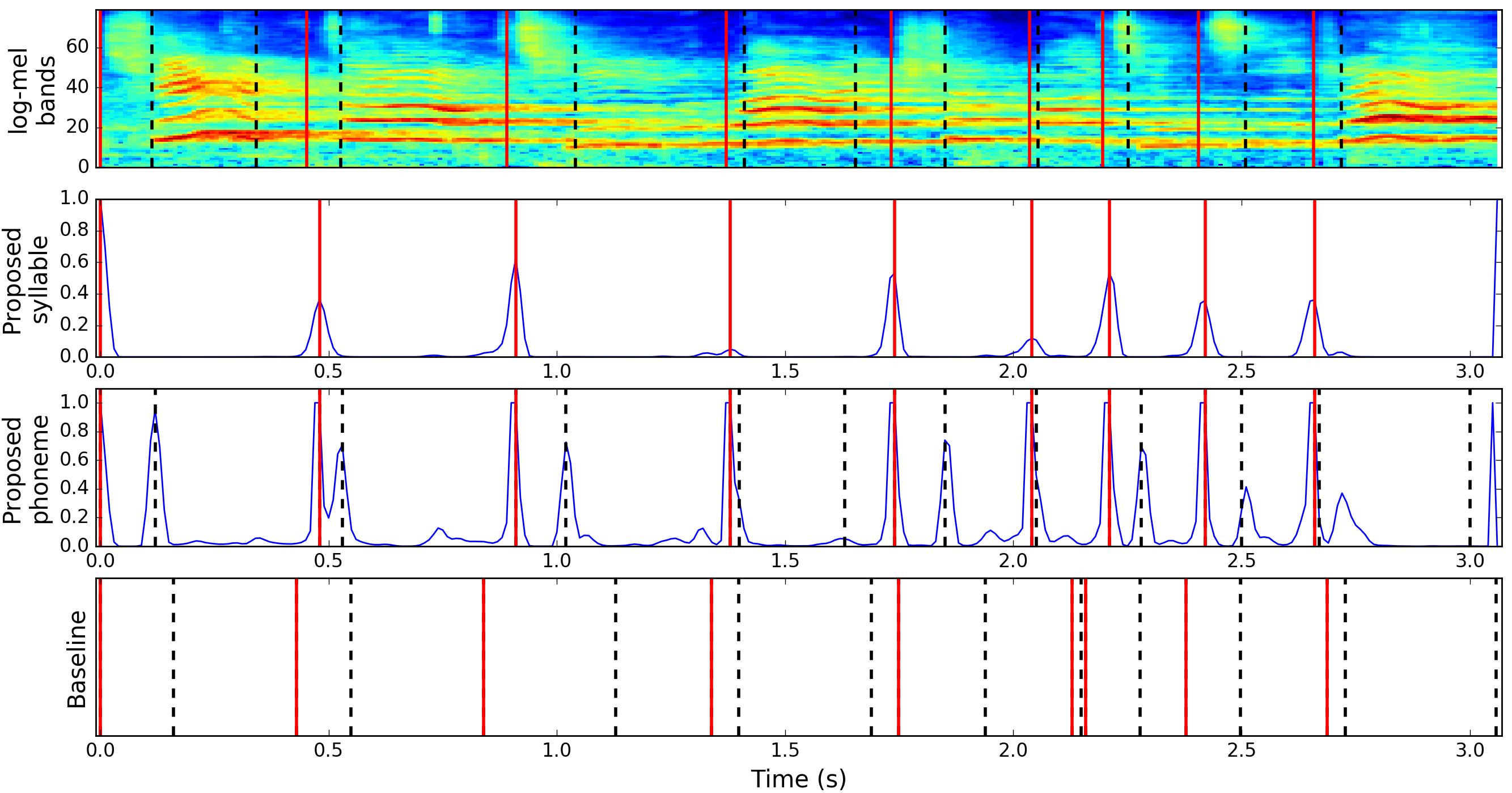}
    \caption{An illustration of the result for a testing singing phrase. The red solid and black dash vertical lines are respectively the syllable and phoneme onset positions (1st row: ground truth, 2nd and 3rd rows: proposed method, 4th row: baseline method). The blue curves in the 2nd and 3rd row are respectively the syllable and phoneme ODFs. }
    \label{fig:results_example}
\end{figure}

Fig \ref{fig:results_example} shows an result example for a testing singing phrase. The syllable/phoneme labels, baseline emission probabilities matrix and alignment path are omitted for the plot clarity, and can be found in the link\textsuperscript{\ref{ft:github}}. Notice that there are some extra or missing onsets in the detection. This is due to the inconsistency between the coarse duration input and the ground truth -- students might add or delete some phonemes in the actual singing. Also notice that in the 3rd row, the two detected phoneme onsets within the last syllable are not in the peak positions of the ODF. This is due to that the onsets is inferred by taking into account both ODF and the \textit{a priori} duration model, and the latter partially constraints the detected onsets.

The biggest advantage of the proposed method is the language-independency, which means that the pre-trained CNN model can be eventually applied to the singing voice of various languages because they could share the similar temporal-spectral patterns of phoneme transitions. Besides, the Viterbi decoding of the proposed method (time complexity $O(TS^2$), $T$: time, $S$: states) is much faster than the HSMM counterpart (time complexity $O(TS^2+T^2S)$). An interactive jupyter notebook demo for showcasing the proposed algorithm is provided for running in Google Colab\footnote{\url{https://goo.gl/BzajRy}}. 

\section{Conclusions}

In this paper, we presented a language-independent singing voice segmentation method by first jointly learning the syllable and phoneme ODFs using a CNN model, then inferring the onsets and segmented labels using a duration-informed HMM. We also presented a jingju solo singing voice dataset with manual boundary and label annotations. For the evaluation, we compared the proposed method with a baseline forced alignment method based on a language-dependent HSMM. The evaluation results showed that the proposed method outperforms the baseline in both segmentation and onset detection tasks. We attribute the performance improvement of the proposed method to the efficient use of the onset and duration information for our relatively small dataset. However, the proposed method is not able to solve the phoneme insertion or deletion problems when there is a mismatch between the prior coarse duration information and the actual singing. We are improving the algorithm to overcome this limitation by using recognition-based methods.

\section{Acknowledgements}

This work is supported by the CompMusic project (ERC grant agreement 267583).

\bibliographystyle{IEEEtran}

\bibliography{mybib}

% \begin{thebibliography}{9}
% \bibitem[1]{Davis80-COP}
%   S.\ B.\ Davis and P.\ Mermelstein,
%   ``Comparison of parametric representation for monosyllabic word recognition in continuously spoken sentences,''
%   \textit{IEEE Transactions on Acoustics, Speech and Signal Processing}, vol.~28, no.~4, pp.~357--366, 1980.
% \bibitem[2]{Rabiner89-ATO}
%   L.\ R.\ Rabiner,
%   ``A tutorial on hidden Markov models and selected applications in speech recognition,''
%   \textit{Proceedings of the IEEE}, vol.~77, no.~2, pp.~257-286, 1989.
% \bibitem[3]{Hastie09-TEO}
%   T.\ Hastie, R.\ Tibshirani, and J.\ Friedman,
%   \textit{The Elements of Statistical Learning -- Data Mining, Inference, and Prediction}.
%   New York: Springer, 2009.
% \bibitem[4]{YourName17-XXX}
%   F.\ Lastname1, F.\ Lastname2, and F.\ Lastname3,
%   ``Title of your INTERSPEECH 2018 publication,''
%   in \textit{Interspeech 2018 -- 19\textsuperscript{th} Annual Conference of the International Speech Communication Association, September 2-6, Hyderabad, India Proceedings, Proceedings}, 2018, pp.~100--104.
% \end{thebibliography}

\end{document}